\documentclass[12pt]{iopart}
\usepackage{graphicx,latexsym,stmaryrd,iopams}

\usepackage{color}

\begin{document}

\title{The $1$D interacting Bose gas in a hard wall box}

\author{M.T. Batchelor$^\dag$, X.W. Guan$^\dag$,  N. Oelkers$^\dag$  and  C. Lee$^\ddag$}
\address{\dag\
{\small Department of Theoretical Physics, Research School of Physical Sciences and Engineering, and}
{\small Department of Mathematics, Mathematical Sciences Institute,}
{\small The Australian National University, Canberra ACT 0200, Australia}}
\address{\ddag\
{\small Nonlinear Physics Centre, Research School of Physical Sciences and Engineering, and
ARC Centre of Excellence for Quantum-Atom Optics, The Australian National University, 
Canberra ACT 0200, Australia
}}


\begin{abstract}
We consider the integrable one-dimensional $\delta$-function interacting Bose gas 
in a hard wall box which is exactly solved via the coordinate Bethe Ansatz. 
The ground state energy, including the surface energy, is derived from the
Lieb-Liniger type integral equations.
The leading and correction terms are obtained in the weak coupling and
strong coupling regimes from both the discrete Bethe equations and the 
integral equations.
This allows the investigation of both finite-size and boundary effects in the
integrable model.
We also study the Luttinger liquid behaviour by calculating Luttinger parameters 
and correlations. 
The hard wall boundary conditions are seen to have a strong effect on the 
ground state energy and phase correlations in the weak coupling regime. 
Enhancement of the local two-body correlations is shown by application
of the Hellmann-Feynman theorem.
\end{abstract}



\maketitle

\section{Introduction}

Quantum Bose and Fermi gases of ultracold atoms continue to attract considerable interest
since the experimental realization of atomic Bose-Einstein condensates
(BEC) \cite{BEC0,BEC1,BEC2,BEC3} and the pair condensation of 
fermionic atoms \cite{BEC-F1,BEC-F2,BEC-F3}. 
Particular attention has been paid to one-dimensional ($1$D) Bose gases, 
which are seen to exhibit the rich and novel effects of quantum many-body systems
\cite{EXPboson2,Bose1,Bose2,Bose3,Bose4,Bose5,Bose6,Bose7}. 
As a consequence, there has been a revival of interest in the exactly
solved $1$D model of interacting bosons. 
It is well known that the $\delta$-function interacting Bose gas is
integrable \cite{LL,MC} and can be realized via short-range interactions with an
effective coupling constant $g_{\rm 1D}$ \cite{Bose1}. 
This constant is determined through an effective $1$D scattering length 
$a_{\rm 1D}\approx a^2_{\perp}/a$, where $a_{\perp}$ is the characteristic
length along the transverse direction and $a$ is the $3$D scattering length. 
The ratio of the average interaction energy to the kinetic energy, 
$\gamma=\frac{mg_{\rm 1D}}{\hbar^2n}$, is used to characterize
the different physical regimes of the $1$D quantum gas. 
Here $m$ is the atomic mass and $n$ is the boson number density. 
In the weak coupling regime, i.e., $\gamma \ll  1$, the wave functions of the 
bosons are coherent.  
In this regime, the density fluctuations are suppressed and the phase correlations 
decay algebraically at low temperatures.
Thus the $1$D Bose gas can undergo a quasi BEC. 
However, in the opposite limit, i.e., the Tonks-Girardeau limit
$\gamma \gg 1$, the bosons behave like impenetrable hard core
particles, the so called Tonks-Girardeau gas \cite{KG}.
In this regime the single-particle wave
functions become decoherent and the system acquires fermionic properties.

\begin{figure}[t]
\begin{center}
\includegraphics[width=.70\linewidth]{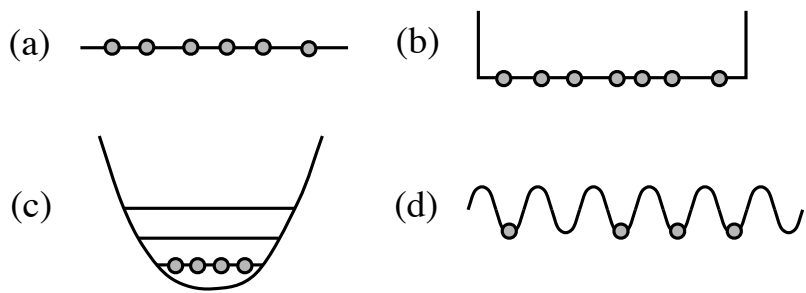}
\end{center}
\caption{
Schematic diagrams of different models used to describe experiments on
quasi-1D bosons with contact interaction:
(a) Bosons on a line with periodic boundary conditions (no potential) -- 
exactly solved via Bethe Ansatz \cite{LL}.
(b) Bosons confined to a hard-wall box -- also exactly solved via Bethe Ansatz (this paper).
Under certain conditions the hard walls can mimic a trapping potential.
(c) Bosons confined by a harmonic trap potential -- not exactly solved, but
closer to experimental conditions.
d) Lattice version realized by, e.g., experiments with optical lattices. 
The corresponding Bose-Hubbard type model is in general not integrable.
}
\label{fig:cartoon}
\end{figure}

The $1$D Bose gas is realized experimentally by tightly confining the atomic cloud in
two (radial) dimensions and weakly confining it along the axial direction. 
The motion along the radial direction is then frozen to zero point
oscillations \cite{Exp-B1,Exp-B2,Exp-B3,Exp-B4,Exp-B5}, making the gas  
effectively one-dimensional.
Anisotropic trapping along the radial and axial directions can form either a 
$2$D optical lattice or $1$D tubes. 
There have been two types of $1$D quantum gases, the
lattice Bose gas \cite{Exp-B1,Exp-B2,Exp-B3} and the continuum Bose gas
confined in a harmonic potential along the axial direction \cite{Exp-B4} (see \fref{fig:cartoon}). 
From a theoretical point of view, the former is usually described
by the Bose-Hubbard model while the latter is described by the $1$D 
interacting Bose gas.
The Bose-Hubbard model is not integrable, except for a special case \cite{ZJMG}, 
which corresponds to two sites.
On the other hand, the
trapping potential along the axial direction breaks the integrability of the 
$1$D Bose gas. 
However, the long-wavelength properties of the $1$D Bose
gas and the Bose-Hubbard model can be described by a Luttinger liquid, 
owing to the universality of the low energy excitations,
i.e., gapless excitations with a linear low-energy
excitation spectrum  and power-law decay in the correlations \cite{Haldane}.

The exactly solved $1$D interacting Bose gas has been
extensively studied \cite{MAT,KOR,Tbook,Thacker,MC,Suth,Wadati,SEN,Amico,BGM}. 
The ground state energy and low energy excitations \cite{LL,L}, 
thermodynamic behaviour \cite{Y-Y}, finite-size effects \cite{F-S-C}, 
correlation functions \cite{KOR,Corre} and Luttinger liquid behaviour
\cite{CAZA1,CAZA2,LUTT1} have been investigated via various methods. 
The signature of the $1$D Bose gas is strongly influenced by
the interaction strength and the external trapping potential. 
The effects of spatial inhomogeneity and finite temperature are other 
considerations to be taken into account under experimental conditions.  
To this end, several approximation schemes have been adopted to 
describe the main features of the $1$D trapped Bose gas.  
In particular, the local density approximation
\cite{Bose3,Oliva,LDA,LDA2} is widely used for calculating the density
profiles of bosons and fermions in harmonic traps.

Now for a finite number of bosons and finite system size, boundary effects are expected
to be pronounced at low temperature \cite{HW1,HW2,HW3}. 
Indeed, significantly different quantum effects should be exhibited by a finite number of bosons 
confined in a finite hard wall box.
For example, in the weak coupling regime, macroscopic states lie on the
zero point oscillations as if the system undergoes BEC. 
The density expectation value exhibits Friedel oscillations and the correlation decay 
is slower than in the periodic case, due to the enhancement of the density and phase stiffness. 
The boundary conditions also have an effect on the phase correlations near the boundaries. 
The ground state of the $1$D interacting Bose gas with hard wall boundary conditions has no
momentum pairing (with a $-k$ for each $k$) compared to the period case, 
because of the missing translational symmetry.
The hard wall boundary conditions have been  experimentally
realized by square potentials with very high barriers \cite{HW-Exp1}. 
Most recently,  BEC have been produced in a novel optical box trap \cite{BECbox}, 
in which atom numbers are as small as $5 \times 10^2$.
More experiments in this direction can be anticipated \cite{CSZ,CB}.
These are our motivations for studying the ground state properties of the $1$D interacting
Bose gas confined in a hard wall box.

The $1$D interacting Bose gas with hard wall boundaries was solved
by Gaudin in the early 1970's \cite{Gaudin1}. 
Gaudin calculated the surface energy via the Bethe Ansatz solution in the thermodynamic limit. 
Very recently, this model was studied via Haldane's harmonic liquid theory \cite{HW3,CAZA2}, 
which describes the long wave-length properties of the $1$D fluid in terms of the density and phase fluctuations. 
The correlation functions of the Tonks-Girardeau gas have also been studied with hard wall boundary
conditions \cite{Forr}.

The paper is organized as follows. 
In Section \ref{BA}, we present the Bethe Ansatz wave functions and Bethe equations for the $1$D
interacting Bose gas with hard wall boundary conditions. 
Details of the Bethe Ansatz solution are given in Appendix A.
In Section \ref{AS}, we derive the ground state energy via asymptotic roots of the Bethe
equations in the strong and weak coupling limits. 
We derive the surface energy through the continuum integral equations in Section \ref{TDL}. 
In Section \ref{WPSEM}, we calculate the ground state energy in the strong
and weak coupling limits using Wadati's power series expansion 
method \cite{Wadati,Wadati-Kato,Wadati-Iida}.
A discussion of the connection between the $1$D Bose gas
trapped by an harmonic potential and the exactly solved model is given in Section \ref{LDA}.
The low-energy properties are discussed in Section \ref{sec:LL},
with concluding remarks given in \sref{sec:con}.

\section{The Bethe Ansatz solution}
\label{BA}

The $1$D quantum gas of $N$ bosons with $\delta$-function interaction in a 
hard wall box of length $L$ is described by the Hamiltonian 
\begin{equation}
{\cal H}=-\frac{\hbar ^2}{2m}\sum_{i = 1}^{N}
\frac{\partial^2}{\partial x_i^2}+\,g_{\rm 1D} \sum_{1\leq i<j\leq N} \delta (x_i-x_j)
\label{Ham-1}
\end{equation}
where the hard walls are defined via the boundary conditions  \cite{Gaudin1}
\begin{equation}
\Psi(x_1=0,x_2,\ldots,x_N) = 0,\,\,\,\,\Psi(x_1,x_2,\ldots,x_N=L) = 0\label{HW-BC}.
\end{equation}
Here $g_{\rm 1D} ={\hbar ^2 c}/{m}$ is an effective $1$D coupling constant with
scattering strength $c$. 
The wavefunction $\Psi$ must be totally symmetric in all its arguments, 
as required for a bosonic system.

For harmonic trapping along the axial direction, the scattering strength is
given by $c=\frac{2}{|a_{\rm 1D}|}$. 
The trapping potential should be added to the Hamiltonian \eref{Ham-1} as an external field.
However, harmonic potentials appear to break the integrability of the model.
Fortunately the integrability of the model is preserved by the hard wall boundary 
conditions \cite{Gaudin1}.
This provides us with an opportunity to study the signature of the $1$D Bose gas in a hard wall
box in an exact fashion. 
For simplicity, we set $\hbar=2m=1$ in the following.

The explicit solution of the model via the coordinate Bethe Ansatz is described in Appendix A.
The wavefunction is given by
\begin{equation}
\fl
\Psi_{\left\{\epsilon_ik_i\right\}}(x_1,\ldots ,x_N)=\sum_{\epsilon_1,\ldots,\epsilon_N}
\sum_{P}\epsilon_1\cdots \epsilon_N A(\epsilon_1 k_{P1}\cdots \epsilon_Nk_{PN}) 
\e^{\mathrm{i}(\epsilon_1k_{P1}x_1+\cdots +\epsilon_Nk_{PN}x_N)} \label{WF}
\end{equation}
where the sum extends over all $N!$ permutations $P$ and all signs $\epsilon_i =\pm$
(see Appendix A).
The wavefunction is valid in the domain $0\le x_1<\ldots< x_N \le L$ and can be continued via
symmetry in all coordinates $x_i$.
The wavefunction coefficients $A(\epsilon_1 k_{Px_1}\cdots \epsilon_Nk_{Px_N})$ are
determined by the Bethe roots, or wave numbers, $k_i$ via
\begin{equation}
\fl
A(\epsilon_1 k_{P1}\cdots \epsilon_Nk_{PN})=(-)^P\prod_{i=1}^{N-1}\prod_{j>i}^N
(\epsilon_1k_{Pi}-\epsilon_jk_{Pj}+\mathrm{i}c)(\epsilon_i k_{Pi}+\epsilon_jk_{Pj}-\mathrm{i}c). \label{coff-A}
\end{equation}
In the above equation $(-)^P$ denotes a $(\pm)$ sign factor associated with even/odd permutations. 
The wave numbers satisfy the Bethe equations
\begin{equation}
\e^{\mathrm{i}2k_jL}=- \prod^N_{\ell = 1}
\frac{(k_j-k_\ell+\mathrm{i}\, c)(k_j+k_\ell+\mathrm{i}\, c)}{(k_j-k_\ell-\mathrm{i}\, c)(k_j+k_\ell-\mathrm{i}\, c)}
\qquad \forall\ j=1,\ldots,N.
\label{BE}
\end{equation}
The energy eigenvalues are as usual given by
\begin{equation}
E=\sum_{j=1}^Nk_j^2.
\end{equation}
Like the periodic boundary condition case  \cite{KOR}, the Bethe roots $k_i$ are known to be real for 
repulsive interactions ($c > 0$), but they may become complex for attractive interactions ($c < 0$).
Here we consider the repulsive regime.
Free bosons are recovered for $c=0$, i.e., 
$k={\pi}n/{L}, n\in \mathbb{N}$ (see Appendix A).

\section{Asymptotic solutions to the Bethe equations}
\label{AS}

In contrast with periodic boundary conditions, the ground state no longer contains $\pm$ momentum 
pairs due to the reflection of quasi momenta at the boundaries. 
As a result the total quasi momentum $\sum_{j=1}^N k_j$ is not conserved in this case.
We first examine the asymptotic solutions of the Bethe equations \eref{BE} in the strong and 
weak coupling limits.

\subsection{Tonks-Girardeau regime}

It is well known that in the strong coupling regime, i.e., $\gamma \gg 1$, the $1$D Bose gas with 
repulsive interaction behaves like a gas of weakly interacting fermions \cite{CB}. 
In the limiting case $c=\infty$ the exact solution for periodic boundary conditions and harmonic 
trapping has been given for impenetrable bosons \cite{KG,Bose6}.
In the hard wall setting the fermionic behaviour can be seen from the ground state energy. 
Define the variables $z_j={L}k_j/{N}$ and $\gamma={L}c/{N}$.
Then for $\gamma\gg 1/N$ the Bethe equations \eref{BE} can be written in the asymptotic form
\begin{eqnarray}
\fl
\exp({\mathrm{i} 2 N z_j})  \approx && 1-2\sum^{N}_{\ell=1}
\left\{\frac{(z_j-z_{\ell})^2}{\gamma^2}+\frac{(z_j+z_{\ell})^2}{\gamma^2}\right\}
-4\sum_{\ell=1}^{N-1}\sum_{\ell ' < \ell }^{N}\frac{(z_j-z_{\ell})}{\gamma}\frac{(z_j+z_{\ell '})}{\gamma}
\nonumber\\
& &
- \,2\,  \mathrm{i} \sum_{\ell=1}^N\left\{\frac{(z_j-z_{\ell})}{\gamma}+\frac{(z_j+z_{\ell})}{\gamma}\right\}
\qquad \forall\ j=1,\ldots,N
\label{APPROX}
\end{eqnarray}
in which the summations exclude $\ell=j$ and $\ell'=j$. 
Here we restrict the solutions to $z_j>0$. 
The asymptotic Bethe roots 
\begin{equation}
z_j \approx \frac{\pi  j}{N}\left({1+\frac{2(N-1)}{N\gamma}}\right)^{-1}
\qquad \forall\ j=1,\ldots,N
\end{equation}
for the ground state energy follow from the condition that the eqns \eref{APPROX} be consistent.
It follows that in this limit the ground state energy per particle is given by
\begin{equation}
\frac{E}{N}\approx \frac{\pi^2}{6 L^2}(N+1)(2N+1)\left(1+\frac{2(N-1)}{Lc}\right)^{-2}.
\label{E-TK-1}
\end{equation}

We emphasize that these asymptotic solutions are very accurate for the Tonks-Girardeau regime. 
We will compare the ground state energy \eref{E-TK-1} with numerical solutions of
the continuum integral equation, which is the hard-wall analogue of the Lieb-Liniger 
integral equation, in Section \ref{TDL}.
The explicit form for the wave numbers $k_j$ also allows an in principle calculation of the asymptotic 
correlation functions directly from the wave function \eref{WF}. 
Switching back to real physical units, the ground state energy per particle \eref{E-TK-1} can
also be written as 
\begin{equation}
\frac{E}{N}\approx \frac{\hbar^2 n^2}{2m}\left({e_0(\gamma)+\frac{1}{N}e_f(\gamma)}\right).
\label{E-TK-2}
\end{equation}
Here the bulk energy $e_0(\gamma)$ and the surface energy $e_{\rm f}(\gamma)$ are given by
\begin{eqnarray}
e_0(\gamma)&=&\frac{\pi^2}{3}\left(1+\frac{2}{\gamma}\right)^{-2}\label{e0}\\
e_{\rm f}(\gamma)&=&\frac{\pi^2}{2}\left(1+\frac{2}{\gamma}\right)^{-2}.\label{ef}
 \end{eqnarray}

\begin{figure}[t]
\vskip 5mm
\begin{center}
\includegraphics[width=.70\linewidth]{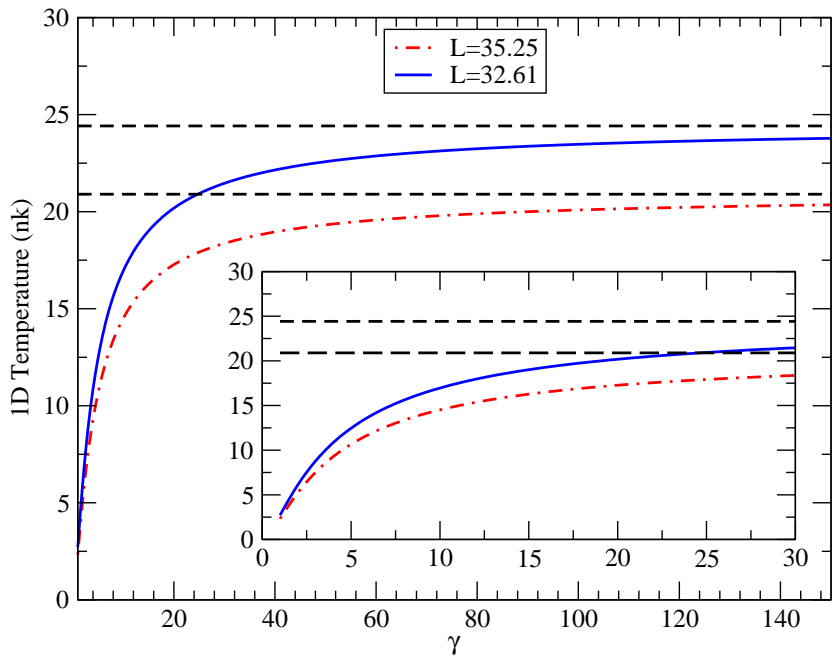}
\end{center}
\caption{The $1$D temperature $T_{\rm 1D}$ obtained from the asymptotic result \eref{E-TK-2} 
versus interaction strength $\gamma$ for $N=37$ ${}^{87}$Rb atoms confined in 1D boxes of 
length $L=35.25 \mu$m and $L=32.61 \mu$m. The asymptotic result \eref{E-TK-2} is valid for a wide range of interaction, i.e. $\gamma\gg 1/N$.  
The horizontal dashed lines are the corresponding temperatures in the Tonks-Girardeau limit.
The inset shows the $1$D temperatures in the regime $1<\gamma <30$. }
\label{fig:1DT}
\end{figure}

A useful quantity is the $1$D temperature $T_{\rm 1D}=\frac{2E}{N k_B}$, which is just the ground state 
energy in different units \cite{Exp-B4}.
We plot $T_{\rm 1D}$ obtained from \eref{E-TK-2} as a function of the interaction strength 
$\gamma$ in \fref{fig:1DT}
for a gas of $N=37$ bosons confined in boxes of length $L=35.25$ $\mu$m and $L=32.61$ $\mu$m. 
These are the same parameters as in Figure $3$ of Ref~\cite{Exp-B4}.
The dashed horizontal lines are the corresponding values of $T_{\rm 1D}$ in the Tonks-Girardeau limit.
We remark that for hard wall boundary conditions, the particle density distribution is rather flat
and homogeneous. 
It can be seen that $T_{\rm 1D}$ increases rapidly as $\gamma$ increases in the weak coupling regime.
It then slowly approaches the Tonks-Girardeau energy as $\gamma$ tends to infinity. 
In an actual experiment, the length of the atomic cloud varies with the interaction strength. 
In that case the $T_{\rm 1D}$ will increase smoothly as $\gamma$ increases. 
Most significantly, $T_{\rm 1D}$ is sensitive to the length of the hard wall box. 
The smaller the length of the box, the larger the `quasi-momentum' of the particles.
We note also that the surface energy is positive.
In the thermodynamic limit, the ground state energy for periodic boundary conditions 
is proportional to the linear density $n$. 
Therefore, in the thermodynamic limit, the ground state energy per particle of the Bose gas 
in a hard wall box can be considered as an excited state of a Bose gas with $2N$ particles 
in a periodic box  of length $2L$ \cite{Gaudin1}. 
We will study the ground state properties for the hard wall box further in Section \ref{TDL}.

\subsection{Weak coupling regime}

The ground state properties in the weak coupling limit are both subtle and interesting. 
So far it has proved difficult to reach the weak coupling regime via anisotropic trapping in 
experiments \cite{Exp-B4} and it is necessary to sharply define the criterion for weak 
coupling \cite{LUTT1,LDA}.  
In the experiment, the $1$D regime is reached if the radial zero point oscillation length 
$l_0=\sqrt{\hbar/(m\omega_{\perp})}$ is much smaller than the axial correlation length 
$l_c=\hbar/\sqrt{m \mu}$, where $\omega_{\perp}$ is the frequency of the radial trap and $\mu$ is
the chemical potential of the $1$D system, thus the condition is $\mu \ll \hbar \omega_{\perp}$ 
for the $1$D trapped system. 
In general, $\gamma \ll1$ at zero temperature is referred to as the weak coupling regime. 
The Thomas-Fermi regime is usually reached for $\mu \gg \hbar \omega$, 
where $\omega$ is the axial oscillation frequency for the harmonic trap. 
In this regime the kinetic energy term can be neglected and the system has a parabolic
density distribution profile \cite{Bose3}, referred to as the Thomas-Fermi BEC. 
However, in the regime $\mu \ll \hbar \omega$ the system is considered
to have a macroscopic occupation in the ground state of the trap with a 
Gaussian density profile \cite{LUTT1}.

For the hard wall boundary conditions, the leading terms in the ground
state energy can be obtained through asymptotic solutions of the Bethe
equations \eref{BE} 
in analogy with the periodic case \cite{BGM}.
Here the wave numbers $k_j$ for the ground state satisfy the algebraic equations
\begin{equation}
k_j = \frac{\pi}{L} + \frac{c}{L} \, {\sum_{\ell=1 \atop \ell\neq j}^{N}}
\left({
        \frac{1}{k_j - k_l } + \frac{1}{k_j + k_l } 
}\right)
\qquad \forall j=1,\ldots,N.
\label{momenta}
\end{equation}
Algebraic equations for periodic boundary conditions have arisen in a number of 
different contexts \cite{BGM}, most notably in the integrable BCS pairing 
models \cite{BCS_review}.
The roots of such equations also describe the 
equilibrium positions of potentials in Calogero systems associated 
with Lie algebras \cite{JPA}.

The solutions of \eref{momenta} give the ground state energy
\begin{equation}
\frac{E}{N}\approx \frac{\pi^2}{L^2}+\frac{3}{2}\frac{(N-1)c}{L}.
\label{MF-weak}
\end{equation}
This is quite different from the periodic boundary case, for which the leading term in 
the ground state energy per particle is $\frac{E}{N}\approx {(N-1)c}/{L}$. 
Here, due to the reflections at the boundaries, the ground state energy for hard wall 
boundary conditions is larger than the energy for periodic boundary conditions. 
For the integrable $1$D Bose gas with periodic boundary conditions, the ground
state energy per particle is known to be given by $\frac{E}{N}\approx \frac{\hbar^2n^2}{2m}e_0(\gamma)$ with 
$e_0(\gamma) \simeq \gamma\left(1-\frac{4}{3\pi}\sqrt{\gamma}\,\right)$ \cite{LL,Tbook,Wadati}.
We argue that this result holds in the regime $1/N^2 \ll \gamma \ll 1$. 
The leading term of the ground state energy per particle for periodic boundary conditions,
i.e. $\frac{E}{N}\approx \frac{\hbar^2n^2}{2m}\gamma $ holds in
the mean-field regime, i.e., $\gamma \ll 1/N^2$, for which the
correction term is proportional to $\gamma^2$ rather than
$\gamma^{3/2}$.
This discrepancy is not totally unexpected, as the Lieb-Liniger integral equation is only valid up to 
terms of order $1/L$.
If the interaction energy is much smaller than the scale of $1/L$, i.e., if $\gamma\ll 1/N^2$, 
results derived from the Lieb-Liniger integral equation are no longer valid in this very weak coupling regime.
On the other hand, in the regime $\gamma \gg 1/N^2$ the zero point oscillation kinetic
energy is much smaller than the interaction energy and is thus negligible --
it is here that one can derive the ground state energy asymptotically from the continuum 
integral equation. 
Finite-size discrepancies between the discrete and integral equation approaches have also been 
noted in Ref.~\cite{SSAC}.

\section{The surface energy}
\label{TDL}

Taking the logarithm on both sides of the Bethe equations \eref{BE} gives
\begin{equation}
\bar{k}_jL=\pi m_j-\sum_{\ell =1}^N\left\{\arctan{\frac{\bar{k}_j-\bar{k}_{\ell}}{c}}+\arctan{\frac{\bar{k}_j+\bar{k}_{\ell}}{c}}\right\}
\label{BE-log}
\end{equation}
where  $j=1,\ldots,N$ and $m_j$ are ordered positive integers, i.e., $1\leq m_1\leq\ldots\leq m_N$. 
Here for later convenience we have denoted the Bethe roots by $\bar k$.
Our calculation takes advantage of the fact that in the thermodynamic limit, the ground state 
energy of the $N$ boson system in a box of length $L$ is equivalent to one half the energy of 
$2N$ bosons with length $2L$ and periodic  boundary conditions, as pointed out by \cite{Gaudin1}. 
It is thus convenient to derive the surface energy from a periodic system of $2N$ bosons with 
a length $2L$, for which the Bethe equations are \cite{LL}\footnote{Of course, one may also 
obtain the same free energy by treating the Bethe equations \eref{BE-log} directly, see, e.g., 
Ref.~\cite{HQB}.}
\begin{equation}
k_jL=2\pi I_j-\sum_{\ell =1}^N\left\{\arctan{\frac{k_j-k_{\ell}}{c}}+
\arctan{\frac{k_j+k_{\ell}}{c}}\right\}\label{BA-PBC}
\end{equation}
for $k_j>0$ and $I_j$ are half-odd integers.  
We now introduce the notation $k_{-j}=-k_j$ and $I_{-j}=-I_j$. 
The difference between $\bar{k}_j$ and $k_j$ can thus be written as
\begin{equation}
(\bar{k}_j-k_j)L=\pi\epsilon_j-\sum_{\ell =-N}^N
\left\{\arctan{\frac{\bar{k}_j-\bar{k}_{\ell}}{c}}-\arctan{\frac{k_j-k_{\ell}}{c}}\right\}
\label{SFE-1}
\end{equation}
where $j=-N,\ldots, N$ and $\epsilon_j$ is a sign factor.

The surface energy is given by
\begin{equation}
E_{\rm f}=\frac{1}{2}\sum_{j=-N}^{N}(\bar{k}^2_j-k^2_j).
\end{equation}
Using $\bar{k}_j-k_j < \pi/L$ and taking the Taylor expansion of equations \eref{SFE-1} 
at $\bar{k}_j=k_j +\Delta k_j$ gives
\begin{equation}
 \Delta k_j L=\pi\epsilon_j-\sum_{\ell=-N}^N\frac{c(\Delta k_j-\Delta k_{\ell})}{c^2+(k_j-k_{\ell})^2}. 
 \label{SFE-2}
\end{equation}
It follows that
\begin{equation}
\fl
\Delta k_j\left(1+\frac{1}{L}\sum_{\ell=-N}^N\frac{c}{c^2+(k_j-k_{\ell})^2} \right)=\frac{1}{L}\left(\pi\epsilon_j+\sum_{\ell=-N}^N\frac{c\Delta k_{\ell}}{c^2+(k_j-k_{\ell})^2}\right).
\end{equation}
Let us  define $k_{j+1}-k_j=\frac{1}{2Lf(k_j)}$, where $f(k)$ is the distribution function \cite{LL},
then the Bethe equations \eref{BA-PBC} become
\begin{equation}
2\pi f(k)=1+2c\int_{-B}^{B}\frac{f(k^{'})}{c^2+(k-k^{'})^2}dk^{'}.
\end{equation}
Here we use the density $n=\frac{2N}{2L}$ and define the cut-off momentum $B$. 
Subsequently, equation \eref{SFE-2} becomes
\begin{equation}
\Delta kf(k)=\frac{1}{2L} \epsilon(k)+\frac{c}{\pi}\int_{-B}^{B}\frac{\Delta k^{'}f(k^{'})}{c^2+(k-k^{'})^2}dk^{'}.
\end{equation}
Here $\epsilon(k) = {\rm sgn}(k)$.
Further defining $ f_{\rm f}(k)=L\Delta kf(k)$, the surface energy is given by
\begin{equation}
E_{\rm f}=\int_{-B}^Bkf_{\rm f}(k)dk
\end{equation}
where $f_{\rm f}(k)$ satisfies the integral equation
\begin{equation}
f_{\rm f}(k)=\frac{1}{2}\epsilon(k)+\frac{c}{\pi}\int_{-B}^{B}\frac{f_{\rm f}(k^{'})}{c^2+(k-k^{'})^2}dk^{'}.
\end{equation}

After the same rescaling as introduced in Ref.~\cite{LL}, i.e., 
$k=Bx,\,c=B\lambda,\, f(Bx)=g(x)$, we find the ground state energy per particle 
to be of the form \eref{E-TK-2}.
The bulk and surface energies are given by
\begin{eqnarray}
e_0(\gamma)&=&\frac{\gamma^3}{\lambda^3}\int^1_{-1}g_0(x)x^2dx\\
e_{\rm f}(\gamma)&=&\frac{\gamma^2}{\lambda^2}\int^1_{-1}g_{\rm f}(x)xdx
\label{LL-E}
\end{eqnarray}
where
\begin{eqnarray}
g_0(x)&=&\frac{1}{2\pi}+\frac{\lambda }{\pi}\int_{-1}^1\frac{g_0(y)}{\lambda ^2+(x-y)^2}dy\label{LL-B}\\
g_{\rm f}(x)&=&\frac{1}{2}\epsilon(x)+\frac{\lambda }{\pi}\int_{-1}^1\frac{g_{\rm f}(y)}{\lambda ^2+(x-y)^2}dy
\label{LL-F}
\end{eqnarray}
with the cut-off condition
\begin{equation}
\gamma \int_{-1}^1g_0(x)dx=\lambda.
\end{equation}

There are various methods which can be used to solve the above equations
\eref{LL-B} and \eref{LL-F}. 
In the next section we derive analytic results from these equations in the strong and weak 
coupling limits.

\section{Application of Wadati's power series expansion method}
\label{WPSEM}

As remarked in references \cite{Gaudin1,Wadati-Iida}, the
Lieb-Liniger integral equation is closely related to the Love
equation for the problem of a circular plate condensator \cite{Sneddon}. 
One can obtain a series expansion for the ground state energy from the integral 
equation \cite{Wadati}.
This method has also been applied to the Yang-Yang integral equations for the thermodynamics 
\cite{Wadati-Kato} and to the Gaudin integral equation for the attractive $\delta$-function interacting 
Fermi gas \cite{Wadati-Iida}.
For the Bose gas with hard walls, Gaudin \cite{Gaudin1} found the leading surface energy term
in the weakly interacting limit. 
In this section, we apply the Wadati method to obtain the leading terms in the ground state energy 
from the integral equations \eref{LL-B} and  \eref{LL-F}.

\subsection{Tonks-Girardeau regime}

In the strong coupling regime the bulk part of the ground state energy per particle 
is given by \cite{LL, Wadati,BGM}
\begin{equation}
e_0(\gamma) \approx \frac{\pi^2}{3}\left(\frac{\gamma}{\gamma+2}\right)^2
\label{wadati}
\end{equation}
which agrees with the asymptotic result  \eref{e0}.
Calculating the first two terms in the expansion
\begin{equation}
g_{\rm f}(x)\approx a_0+a_1x^2
\end{equation}
for the distribution function \eref{LL-F}, we find
\begin{eqnarray}
a_0&=&\frac{\epsilon(x)(\gamma+2)}{2\gamma}\left[1-\frac{2x^2}{3\gamma(\gamma+2)^2}\right]\\
a_1&=&\frac{\epsilon(x)\pi^2}{\gamma(\gamma+2)}.
\end{eqnarray}
This leads to the surface energy 
\begin{equation}
e_{\rm f}\approx \frac{\pi^2}{2}\left(\frac{\gamma}{\gamma+2}\right)
\left(1+\frac{\pi^2(\gamma-2)}{3\gamma (\gamma+2)}-\frac{2\pi^2}{3\gamma^2 (\gamma+2)}\right).
\label{SE-Wada}
\end{equation}
We see that the constant term in the surface energy \eref{SE-Wada} is the same as in \eref{ef}, 
but the leading correction term differs. 
Again this is because the continuum integral equation derived from the Bethe equations is only valid 
for terms of order up to $1/L$.

\subsection{Weak coupling regime}

In this regime the leading terms of the distribution function are found to be
\begin{equation}
g_{\rm f}(x)\approx \frac{\epsilon(x)}{\sqrt \gamma}\sqrt{1-x^2}.
\end{equation}
The surface energy $e_{\rm f}\approx \frac{8}{3}\sqrt{\gamma}$ follows from 
equation \eref{LL-F}.
Here we keep only the leading term, as for a
large number of particles, the contribution from other terms is negligible. 
The ground state energy per particle in the regime $1/N^2 \ll \gamma \ll 1$ is again of
the form  \eref{E-TK-2}
where the leading bulk and surface energy terms are 
\begin{eqnarray}
e_0(\gamma)&\approx &\gamma(1-\frac{4}{3\pi}\sqrt{\gamma})\label{e0-weak}\\
e_{\rm f}(\gamma)&\approx & \frac{8\sqrt{\gamma}}{3}.\label{ef-weak}
\end{eqnarray}

\begin{figure}[t]
\vskip 10mm
\begin{center}
\includegraphics[width=.70\linewidth]{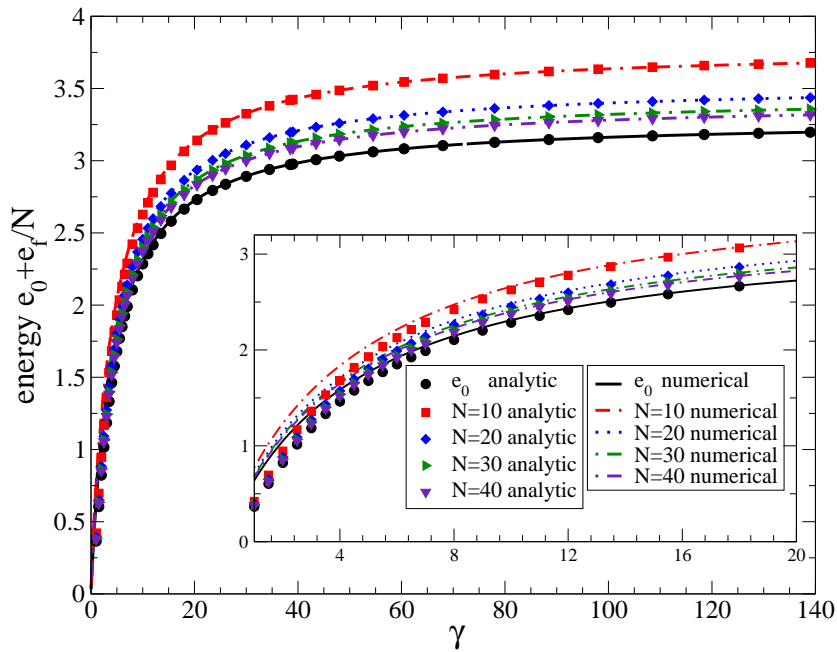}
\end{center}
\caption{The ground state energy $e_0+\frac{1}{N}e_{\rm f}$ versus the
interaction strength $\gamma$ for $N=10,20,30,40$ particles.
For each size there is a comparison between 
the numerical solution of the integral equations \eref{LL-B} and \eref{LL-F} and the analytic 
expression for strong coupling \eref{E-TK-2}, with \eref{e0} and \eref{ef},  derived from the 
discrete Bethe equations \eref{BE}.  
A generally good agreement between the numerical and
analytic results is visible.
The lowest curve is obtained in the thermodynamic limit.}
\label{fig:e0f-strong}
\end{figure}

\begin{figure}[t]
\vskip 10mm
\begin{center}
\includegraphics[width=.80\linewidth]{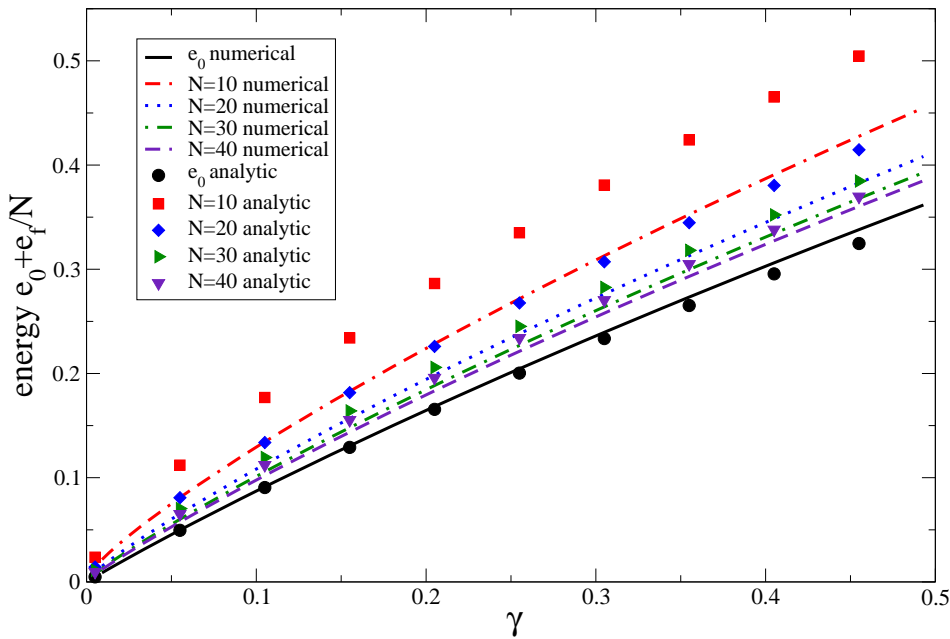}
\end{center}
\caption{
Ground state energy $e_0+\frac{1}{N}e_{\rm f}$ in the weak coupling regime
 versus the interaction strength $\gamma$ for $N=10,20,30,40$.
 For each size there is a comparison
between the numerical result evaluated from the integral equations \eref{LL-B} and \eref{LL-F} and
the analytic expressions for weak coupling, \eref{e0-weak} and \eref{ef-weak}. 
A slight discrepancy between the numerical and analytic results appears for weak coupling. 
This discrepancy becomes small if the particle number is very large. 
The analytic expressions are expected to hold in the region ${1}/{N^2}\ll \gamma  \ll 1$.}
\label{fig:e0f-weak}
\end{figure}

So far we have derived some analytic results for the ground state
energy of the interacting $1$D Bose gas with hard wall boundary conditions. 
One may also perform direct numerical calculations using the integral equations 
\eref{LL-B} and \eref{LL-F}, as originally done in the bulk \cite{LL}.
In doing this we see that the ground state energy \eref{E-TK-2}, with \eref{e0} and \eref{ef},
is consistent with the result obtained from the the integral equations \eref{LL-B} and \eref{LL-F} 
for $\gamma> 1$, with best agreement found for $\gamma> 5$ (see \fref{fig:e0f-strong}). 
A comparison between the analytic result and numerical calculation
for weak coupling is presented in Figure \ref{fig:e0f-weak}.  
A discrepancy between the numerical and analytic results is
evident for weak coupling. 
This implies that the next leading term
in the surface energy \eref{ef-weak} is necessary.
As expected, there is a difference between the finite-size results and the limiting curve
obtained in the thermodynamic limit.


\section{Comparison with the $1$D Bose gas trapped by harmonic potentials}
\label{LDA}

The experimental realization of the Tonks-Girardeau gas trapped
in harmonic potentials has involved the measurement of momentum
distribution profiles \cite{Exp-B2,Exp-B3}, the ground state energy \cite{Exp-B4} 
and collective oscillations \cite{Exp-B5}.
To model the experiments more closely it is desirable to take into account the
 `soft' boundaries of the harmonic potential rather than the commonly used `hard' 
boundaries of a box.
Unfortunately the axial trapping potential breaks the homogeneity of the integrable model. 
However, if the density varies smoothly in a small interval the systems under consideration can be
thought of as a uniform Bose gas in each small interval \cite{Bose3,LUTT1,LDA}. 
This quasiclassical approach is called the {\sl local density approximation} and is used to 
study the density distribution profile in cases where the chemical potential is much larger
than the level spacing $\hbar \omega$ in the $1$D direction. 
For the equilibrium state the chemical potential of the system in a harmonic trap 
can be taken to be constant.

Applying the local density approximation to the $1$D Bose
gas with periodic boundary conditions at zero temperature, we have
\begin{equation}
\mu (n(z))+V(z)=\mu_0
\end{equation}
where $\mu (n(z))$ is the local chemical potential of the uniform system
and $V(z)=\frac{1}{2}m\omega ^2z^2$ is the local trapping potential. 
We make the Ansatz
\begin{equation}
\left\{ \begin{array}{rr}
\mu (n(z))+V(z)=\mu_0,&\,\,\,\,{\rm for}\,\,|z|\leq R\\
n(z)=0,&\,\,\,\,{\rm for}\,\,|z|> R
\end{array}\right.
\end{equation}
where $R$ is the atomic cloud radius given by $R=\sqrt{\frac{2\mu_0}{m\omega^2}}$. 
The density profile of the system can be obtained by using the normalization condition
\begin{equation}
\int_{-R}^{R}n(z)dz =N.
\end{equation}
With help of the analytical expressions for the ground state energy of the interacting Bose gas 
it is now straightforward to derive the density profiles in the Thomas-Fermi and Tonks-Girardeau 
regimes.

For the Thomas-Fermi regime, the energy per particle is given by $E_0=\frac{\hbar^2}{2m}n(z)c$
and thus 
\begin{equation}
n(z)=n^0_{\rm TF}\left(1-\frac{z^2}{R_{\rm TF}^2} \right)
\end{equation}
with central density and Thomas-Fermi radius
\begin{eqnarray}
n^0_{\rm TF}=\left(\frac{9m^2\omega^2N^2}{32c\hbar^2}\right)^{\frac{1}{3}},\qquad R_{\rm TF}=\left(\frac{3N\hbar^2c}{2m^2\omega^2}\right)^{\frac{1}{3}}.
\end{eqnarray}
Here $c=2/|a_{\rm 1D}|$. 
The average energy per particle is given by
\begin{equation}
E_{\rm TF}\approx \frac{1}{N}\int_{-R_{\rm TF}}^{R_{\rm TF}}n(z)E_0(n(z))dz=
\frac{1}{5}\left(\frac{9N^2\omega^2\hbar^4c^2}{4m}\right)^{\frac{1}{3}}.
\end{equation}

In the Tonks-Girardeau limit, the local chemical potential is $\mu(n(z))=\frac{\hbar^2\pi^2}{2m}n^2(z)$
and the density distribution is given by $n(z)=n^0_{\rm TG}\sqrt{1-\frac{z^2}{R_{\rm TG}^2}}$, with
\begin{equation}
n^0_{\rm TG}=\sqrt{\frac{2Nm\omega}{\pi^2 \hbar}},\qquad R_{\rm TG}=\sqrt{\frac{2\hbar N}{m\omega }}.
\end{equation}
The average energy per particle in the Tonks-Girardeau limit is 
$E_{\rm TG} \approx \frac14 {\hbar N\omega}$. 
The density profile has been studied in the whole regime \cite{Bose3}. 
The cloud size expands as the interaction strength inceases. 
In the Tonks-Girareau regime, the interaction-dependent radius is given approximately by
\begin{equation}
R\approx R_{\rm TG}\left(1-\frac{32}{9\pi^2}\frac{R_{\rm TG}|a_{\rm 1D}|}{R_0^2}\right).
\end{equation}
Indeed this would slow down the increasing of the average energy with increasing interaction strength 
in the weak coupling regime
if we consider the length of the hard wall box varying with $\gamma$ via the relation $L=2R$.
However, further refinements are necessary for finite systems, i.e., for a finite number of confined bosons.

In the previous sections we have discussed in detail the derivation of
the ground state energy of the $1$D interacting Bose gas confined in a hard wall box.
This integrable system is much easier to treat theoretically than the system 
with harmonic trapping. 
The experimentally measured $1$D energy has been compared with 
theoretical curves obtained using the local density approximation  \cite{Exp-B4}.
However, the theoretical predictions are not convincing for a number of reasons.
First it is not clear if the quantity $\gamma _{\rm avg}$ presented
in the Figures of Ref.~\cite{Exp-B4} corresponds to the dimensionless interacting strength 
$\gamma$ in the uniform Hamiltonian \eref{Ham-1}. 
Secondly, the interaction strength region measured, up to
$\gamma _{\rm avg} < 6$, may be too small to be sure that the Tonks-Girardeau 
regime has been reached.
Our theoretical results indicate that finite-size effects induced from the
number of particles, the system size and the boundary conditions are not
negligible in the weak coupling and Tonks-Girardeau regimes.

There is  some similarity between harmonic trapping and hard wall box confinement. 
For $\gamma=0$, the kinetic zero point oscillation energy is $\frac{1}{4}\hbar \omega$ 
for axial harmonic trapping.
If we confine the $1$D Bose gas in a hard wall box of length $L=2R_0$, where 
$R_0=\sqrt{\frac{2\hbar }{m\omega}}$ is the characteristic length of the harmonic oscillator, 
the kinetic energy per particle is $\frac{\pi^2}{16}\hbar \omega$ for the hard wall
boundary conditions, which is much larger than the kinetic zero point 
oscillation energy, for harmonic trapping.
For the Tonks-Girardeau regime, if we confine the $1$D
Bose gas in a length $L=2R_{\rm TG}$, the $1$D energy per particle is
$\frac{\pi^2}{16}(\frac{1}{3}+\frac{1}{2N})N\hbar \omega$, which is
almost the same as the average $1$D energy $E_{\rm TG}\approx \frac14 {\hbar N\omega}$
 for harmonic trapping.

The boundary effects are more pronounced in the weak coupling limit.
If one takes the same zero point kinetic energy and
$1$D ground state energy for the hard wall box as that for harmonic trapping, 
the size of the hard wall box for the $\gamma=0$ and $\gamma=\infty$ limits should be 
$L_0=\sqrt{\frac{2\hbar\pi^2}{m\omega}}$ and 
$L_{\rm TG}=\sqrt{\frac{2\hbar N\pi^2}{m\omega }(\frac{1}{3}+\frac{1}{2N})}$, respectively. 
Recall that we plotted the $1$D temperature as a function of the interaction strength
for the hard wall boundary conditions with $L_{\rm TG}=32.61 \mu$m and 
$L=2 R_{\rm TG}=35.25 \mu$m in \fref{fig:1DT}.
These numerical values follow on inserting the physical
parameters for $^{\rm 87}$Rb atoms into the above results, with $N=37$, 
$^{\rm 87}m = 0.1454 \times 10^{-24}$ and $\omega=2\pi \times 27.5$.
In this way \fref{fig:1DT} can be compared with the experimental data in  
Figures $3A$ and 4 of Ref~\cite{Exp-B4}.
As observed in Ref.~\cite{Exp-B4}, the radius $R$ of the atomic cloud 
for harmonic trapping
increases rapidly with the interaction strength in the weak coupling regime,
causing the average energy to increase rather slowly in comparison 
with the hard wall case.

\section{Luttinger liquid behaviour}
\label{sec:LL}

Many one-dimensional models behave like Tomonaga-Luttinger liquids 
due to the universality of the dispersion relation and correlation behaviour. 
The low energy properties are characterized by power-law decay in the correlation functions 
with gapless excitations. 
A universal description of the low-energy properties of one-dimensional interacting systems 
has been given in terms of harmonic liquid theory \cite{Haldane}. 
The $1$D Bose and Fermi gases are included in this theory. 
The Luttinger liquid behaviour of the $1$D Bose gas has recently been studied 
with hard wall boundary conditions \cite{HW3}.
It was found that the particle density exhibits Friedel oscillations with respect to the distance 
to the boundaries. 
The phase and density correlations are influenced by the hard wall boundary effects.
Here we examine this behaviour in the context of the integable model.

\subsection{The Luttinger parameters}

The harmonic liquid approach to the low energy excitations of the $1$D Bose gas 
is described by the effective Hamiltonian \cite{Haldane,CAZA2,HW3}
\begin{equation}
H_{\rm eff}=\case12 \, {\hbar v_s} \int_0^Ldx\left[\frac{\pi}{K}\Pi^2(x)+
\frac{K}{\pi}\Big(\partial\phi(x)\Big)^2\right]
\label{Ham-LL}
\end{equation}
where $K=\sqrt{v_J/v_N}$, with $v_s = \sqrt{v_N v_J}$, is the Luttinger liquid parameter.
Here $v_J $ is the phase stiffness, $v_N$ is the density stiffness and $v_s$ is the 
sound velocity.
For the long-wavelength density fluctuations the density
$\rho(x)=\rho_0+\Pi(x)$ has small deviations from the ground state density $\rho_0$. 
The boson field operator is defined as 
$\Psi^{\dagger}(x)=\sqrt{\rho(x)} \e^{-\mathrm{i}\phi(x)}$, where
$\left[\rho(x),e^{-\mathrm{i}\phi(x')}\right]=\delta(x-x') \e^{-\mathrm{i}\phi(x)}$.
The effective Hamiltonian \eref{Ham-LL} is reduced to the quantum hydrodynamic 
Hamiltonian \cite{CAZA2,HW3}
\begin{equation}
H_{\rm eff}=\sum_{q>0}\hbar\omega(q)b^{\dagger}(q)b(q)+
\frac{\hbar\pi v_s}{2LK}(N-N_0)^2 \label{Ham-LL-Q}
\end{equation}
with regard to the particle-hole excitation modes.
Here $\omega(q)=v_sq$ for $q \ll \rho_0$ and the wave number is restricted to $q>0$. 
$N$ is the total number of particles in the system and $N_0$ is the number in the ground state,
with $ b^{\dagger}(q)$ the creation operator of elementary excitations. 
The low energy excitations are well described by the effective Hamiltonian \eref{Ham-LL-Q} 
with Luttinger parameters $v_s$ and $K$. 
We now study the effect of the hard walls on the Luttinger liquid parameters.
The density stiffness and sound velocity can be derived from the ground state 
energy via the relations \cite{L,CAZA2}
\begin{equation}
v_N=\frac{L}{\pi\hbar }\left[\frac{\partial ^2 E}{\partial N^2}\right],\qquad
v_s=\sqrt{\frac{L}{mn}\left[\frac{\partial ^2 E}{\partial L^2}\right]}.
\end{equation}

\begin{figure}
\vskip 10mm
\begin{center}
\includegraphics[width=.70\linewidth]{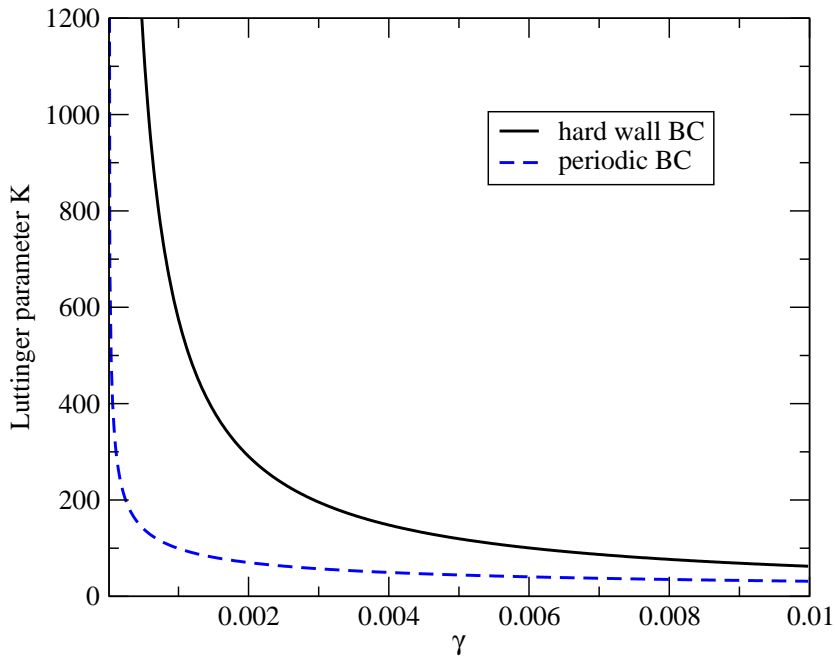}
\end{center}
\caption{Plot of the Luttinger liquid parameter $K$ as a function of the
interaction strength $\gamma$ for $N=20$ bosons with
hard wall boundaries (solid line) and periodic boundary conditions (dashed line).}
\label{fig:K1}
\end{figure}

Although the regime $\gamma \ll 1/N^2$ is difficult to achieve in experiments, the
boundaries nevertheless have a significant effect on the Luttinger behaviour in this regime.
Using the ground state energy \eref{MF-weak}, we have  
\begin{equation}
v_N=\frac{3\gamma}{2\pi^2}v_F, \qquad 
v_s=v_F\sqrt{3\left(\frac{1}{N^2}+\frac{\gamma}{2\pi^2}\right)}. 
\end{equation}
Here the Fermi velocity $v_F={\hbar \pi n}/{m}$. 
We see that the Luttinger liquid parameter 
$K \approx \frac{2\pi ^2}{\sqrt{3}\gamma}\sqrt{\frac{1}{N^2}+\frac{\gamma}{2\pi ^2}}$ tends
to infinity for $\gamma \rightarrow 0$ at a faster rate than for periodic boundaries, for which
$K\approx {\pi}/{\sqrt{\gamma}}$ \cite{CAZA2} (see Figure \ref{fig:K1}). 
The enhancement of $K$ for hard wall boundary conditions leads to a very slow decay of the
phase correlation \cite{CAZA2,HW3}. 
The momentum distribution exponentially decays as $n(p) \sim p^{-\beta}$, where
$\beta=1-\frac{1}{2K}$ \cite{HW3}. 
The system behaves like a BEC in this regime.

\begin{figure}
\vskip 10mm
\begin{center}
\includegraphics[width=.70\linewidth]{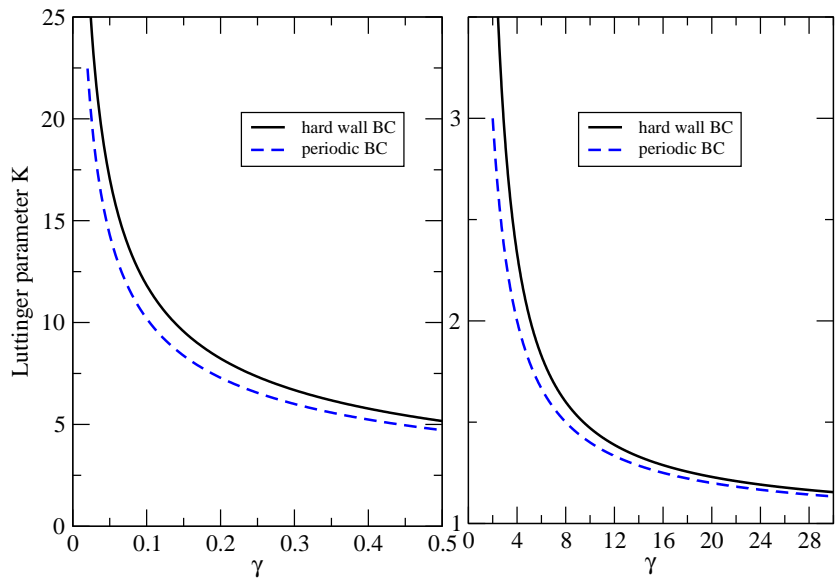}
\end{center}
\caption{Plot of the Luttinger liquid parameter $K$ versus the
interaction strength $\gamma$ for $N=20$ bosons with 
hard wall boundaries (solid line) and periodic boundary conditions (dashed line) in (left) the 
Thomas-Fermi regime  and  (right) the strong coupling regime. 
In the Tonks-Girardeau limit $K$ tends to the value $1+\frac{1}{4N}$.}
\label{fig:WS}
\end{figure}

In the Thomas-Fermi regime $1/N^2\ll \gamma \ll 1$  we use the ground state energy 
in Section 5.2 to obtain the parameters
\begin{eqnarray}
v_N &=&v_F\frac{\gamma}{\pi^2}\left(1-\frac{\sqrt{\gamma}}{2\pi}+\frac{1}{N\sqrt{\gamma}}\right),\\
v_s&=&v_F\frac{\sqrt{\gamma}}{\pi}\left(1-\frac{\sqrt{\gamma}}{2\pi}+\frac{5}{N\sqrt{\gamma}}\right),\\
K&=&\frac{\pi}{\sqrt{\gamma}}
\frac{\sqrt{1-\frac{\sqrt{\gamma}}{2\pi}+\frac{5}{N\sqrt{\gamma}}}}
{1-\frac{\sqrt{\gamma}}{2\pi}+\frac{1}{N\sqrt{\gamma}}}.
\end{eqnarray}
In the strong coupling regime we use the result \eref{E-TK-1} to obtain
\begin{eqnarray}
v_N &=&v_F\left(1+\frac{2}{\gamma}\right)^{-4}\left[1+\frac{1}{2N}\left(1-\frac{4}{\gamma}\right)\right],\\
v_s&=&v_F\frac{\sqrt{1+\frac{3}{2N}}}{\left(1+\frac{2}{\gamma}\right)^2},\\
K&=&\left(1+\frac{2}{\gamma}\right)^{\!\!2}\frac{\sqrt{1+\frac{3}{2N}}}{1+\frac{1}{2N}
\left(1-\frac{4}{\gamma}\right)}.
\end{eqnarray}
We show a comparison of the Luttinger parameter $K$ for the different boundary conditions
in the Thomas-Fermi regime and the strong coupling regime in Figure \ref{fig:WS}. 
The parameter $K$ varies from infinity to $1+\frac{1}{4N}$. 
For weak coupling, $K$ increases more quickly than in the periodic case. 
In the strong coupling limit, $K$ tends to $1$ for both cases.

The density fluctuations are suppressed in the weak coupling limit, 
while they are enhanced in the strong coupling limit. 
This can be seen directly from the leading term of the density expectation value \cite{HW3}
\begin{equation}
\langle\rho(x)\rangle \approx n\left[1-\frac{1}{\pi}\left(\frac{\pi}{n2L|\sin\frac{2\pi x}{2L}|}\right)^{\!\!\!K}
\,\right]\sin(2 \pi nx)
\label{dev}
\end{equation}
with $a\ll x\ll L-a$, where $a$ is the cut-off length to the boundaries. 
For weak coupling we find $a\approx \frac{1}{n\sqrt{\gamma}}$ while in the strong coupling
limit $a\approx \frac{2}{\pi n}$. 
In the weak coupling regime, the density fluctuations are suppressed due to the coherence of the
wave functions.
However, in the strong coupling limit, density fluctuations are evident due to the decoherence of
the wave functions.
These effects can be seen directly from equation \eref{dev}.

\subsection{Local correlation $g_2$}

\begin{figure}
\vskip 10mm
\begin{center}
\includegraphics[width=.80\linewidth]{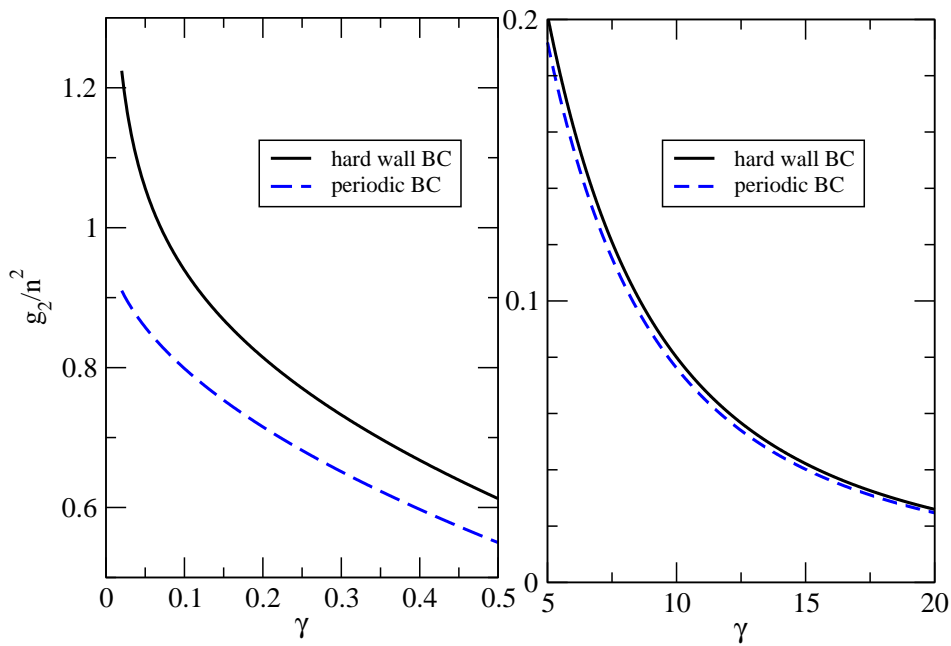}
\end{center}
\caption{The normalized local correlation function $g_2$ versus
coupling strength $\gamma$ for (left) weak coupling and (right) strong
coupling for $N=30$ bosons. 
The enhancement of the correlation function by the hard walls is largest for weak coupling.}
\label{fig:g2}
\end{figure}

It is well known that the local correlations in the $1$D Bose gas decay algebraically \cite{Corre,LUTT1}. 
As for the periodic case, we can calculate the two-body correlation functions through the ground 
state energy for the hard wall boundary conditions. 
Using the Hellmann-Feynman theorem, the $g_2(\gamma)$ correlation function is given 
by \cite{CAZA2, LDA}
\begin{equation}
g_2(\gamma)=n^2\frac{\partial }{\partial \gamma}\left(e_0(\gamma)+\frac{1}{N} 
e_{\rm f}(\gamma)\right).
\end{equation}
We thus obtain the correlation function $g_2$ in the various regions as
\begin{equation}
g_2(\gamma)=\left\{
\begin{array}{ll}
\frac{4\pi^2n^2}{3\gamma^2\left(1+\frac{2}{\gamma}\right)^{\!3}}\left(1+\frac{3}{2N}\right),
&\qquad{\rm for}\,\,\gamma>5\\
n^2\left(1-\frac{2\sqrt{\gamma}}{\pi}+\frac{4}{3N\sqrt{\gamma}}\right),
&\qquad{\rm for}\,\,\frac{1}{N^2}\ll \gamma\ll 1\\
\frac{3}{2}n^2,
&\qquad{\rm for}\,\,\gamma\ll \frac{1}{N^2}.
\end{array}
\right.
\end{equation}
These results are to be compared with the periodic case, for which
\begin{equation}
g_2(\gamma)=\left\{
\begin{array}{ll}
\frac{4\pi^2n^2}{3\gamma^2\left(1+\frac{2}{\gamma}\right)^{\!3}},&\qquad\ \quad\qquad{\rm for}\,\,\gamma>5\\
n^2\left(1-\frac{2\sqrt{\gamma}}{\pi}\right),&\qquad\ \quad\qquad{\rm for}\frac{1}{N^2}\ll \gamma\ll 1\\
n^2,&\qquad\qquad\ \quad{\rm for}\,\,\gamma\ll \frac{1}{N^2}.
\end{array}
\right.
\end{equation}
The enhancement of the correlation function $g_2$ by the hard walls is largest for weak coupling,
as can be seen in Figure \ref{fig:g2}.
This is due to backward scattering increasing the probability of two particles scattering
in comparson with only forward scattering for the periodic case.

\section{Conclusion}
\label{sec:con}

We have considered the integrable interacting $1$D Bose gas in a hard wall box.
The exact Bethe Ansatz solution for the wavefunctions and eigenspectrum 
has been outlined in Appendix A.
The ground state energy, including the bulk and surface energy have been derived from 
the Bethe equations \eref{BE} in different regimes.
For $N$ bosons, these are 
(i) the mean-field regime $\gamma \ll 1/N^2$, 
where the interaction energy is much smaller than the kinetic energy, 
(ii) the Thomas-Fermi regime $1/N^2 \ll \gamma \ll 1$, and
(iii) the strongly interacting Tonks-Girardeau regime $\gamma \gg 1$.
These results have been compared with the ground state energy  obtained from the
continuum Lieb-Liniger-type integral equations in the thermodynamic limit.
The latter results, \eref{LL-E}-\eref{LL-F}, are in agreement with those found by 
Gaudin \cite{Gaudin1}.
The emphasis of our approach has been on finite systems and the effects of the
hard wall boundary conditions. 
It is seen that the finite-size results compare well with those from the continuum integral 
equation, with the exception of the mean-field regime, where the integral equation is not
expected to hold.

A connection to the $1$D Bose gas trapped by  a harmonic potential has also been made. 
The Luttinger liquid parameters, such as the density stiffness, sound velocity,
and the local correlation function $g_2$ have been calculated from the ground state energy
in the various regimes.
It is clearly seen that the hard wall boundary conditions have a larger influence on the phase correlations 
in the weak coupling limit. 
The enhancement of the Luttinger liquid parameter $K$ strongly suppresses
the fluctuation in the ground state density expectation value.
The local correlation $g_2$ is enhanced by the hard wall boundary conditions. 
A significant effect of the hard wall boundary conditions is that the wave-like properties of the 
bosons become more pronounced in the weak coupling regime.
Significantly, the $1$D interacting Bose gas confined in a hard wall box can be experimentally realized.
Future experiments, highlighting the subtle interplay between system size and boundary effects
in ultracold quantum gases, are eagerly awaited. 

\ack

This work has been supported by the Australian Research Council.



\appendix

\section{The Bethe Ansatz solution}

In this Appendix we present the coordinate Bethe Ansatz solution of the
1D interacting Bose gas with hard wall boundary conditions. 
The general form of the wavefunction is that of the spin-$\frac12$ Heisenberg chain with 
open boundaries and additional surface terms \cite{ABBBQ}.

First we consider the standard one particle in a box case, $N=1$, for which the wave function is
\begin{equation}
\Psi(x)=A(k)\e^{{\mathrm i}kx}-A(-k)\e^{-{\mathrm i}kx}.
\end{equation}
From the hard wall boundary conditions \eref{HW-BC}, we have
$A(k)=A(-k)$ and $\e^{{\mathrm i}2kL}=1$. 
The energy is $E=k^2$. 
Thus, up to an overall constant, the wave function is $\Psi(x)=A(k) \sin kx$, with quasi momentum
$k={n\pi}/{L}$, where $n$ is a non-zero integer. 
This describes a trapped particle which has a discrete energy spectrum. 
If $L$ is infinitely large, the particle becomes free and the enery levels will be continuous.

Next we consider $N=2$, with wave function Ansatz
\begin{eqnarray}
\Psi(x_1,x_2)&=&A(k_1,k_2) \,
\e^{\mathrm{i}(k_1x_1+k_2x_2)}+A(k_2,k_1) \, \e^{\mathrm{i}(k_2x_1+k_1x_2)}
\nonumber\\
& &-A(-k_1,k_2) \, \e^{\mathrm{i}(-k_1x_1+k_2x_2)}-A(-k_2,k_1) \, 
\e^{\mathrm{i}(-k_2x_1+k_1x_2)}\nonumber\\
& &-A(k_1,-k_2) \, \e^{\mathrm{i}(k_1x_1-k_2x_2)}-A(k_2,-k_1) \,
\e^{\mathrm{i}(k_2x_1-k_1x_2)}\nonumber\\
& &+A(-k_1,-k_2) \, \e^{-\mathrm{i}(k_1x_1+k_2x_2)}+A(-k_2,-k_1) \, 
\e^{-\mathrm{i}(k_2x_1+k_1x_2)}
\end{eqnarray}
with the domain $0\leq x_1 < x_2 \leq L$.
For $x_1 \neq x_2$, $H \Psi(x_1,x_2)=(k_1^2+k_2^2)\Psi(x_1,x_2)$. 
For $x_1 = x_2$, the consistency condition for the wave function to be continous is
\begin{equation}
\left(\frac{\partial}{\partial x_2}-\frac{\partial}{\partial x_1}\right)\Psi(x_1,x_2)|_{x_1=x_2}=c\Psi(x_1,x_2)
\end{equation}
which leads to the relations
\begin{eqnarray}
&&A(k_1,k_2)=\frac{k_1-k_2+\mathrm{i}c}{k_1-k_2-\mathrm{i}c}A(k_2,k_1),\nonumber\\
&&A(-k_1,k_2)=\frac{k_1+k_2-\mathrm{i}c}{k_1+k_2+\mathrm{i}c}A(k_2,-k_1),\nonumber\\
&&A(k_1,-k_2)=\frac{k_1+k_2+\mathrm{i}c}{k_1+k_2-\mathrm{i}c}A(-k_2,k_1),\nonumber\\
&&A(-k_1,-k_2)=\frac{k_1-k_2-\mathrm{i}c}{k_1-k_2+\mathrm{i}c}A(-k_2,-k_1),
\end{eqnarray}
between the coefficients $A(k_{P1},k_{P2})$.
Using the hard wall boundary conditions \eref{HW-BC}, the unnormalized  wave function is
\begin{eqnarray}
\Psi(x_1,x_2)&=&(k_1-k_2+\mathrm{i}c)(k_1+k_2-\mathrm{i}c) \, \e^{\mathrm{i}(k_1x_1+k_2x_2)}\nonumber\\
& & +(k_1-k_2-\mathrm{i}c)(k_1+k_2-\mathrm{i}c) \, \e^{\mathrm{i}(k_2x_1+k_1x_2)}\nonumber\\
& &-(k_1-k_2+\mathrm{i}c)(k_1+k_2-\mathrm{i}c) \, \e^{\mathrm{i}(-k_1x_1+k_2x_2)}
\nonumber\\
& &-(k_1-k_2+\mathrm{i}c)(k_1+k_2+\mathrm{i}c) \, \e^{\mathrm{i}(k_2x_1-k_1x_2)}
\nonumber\\
& &-(k_1+k_2+\mathrm{i}c)(k_1-k_2-\mathrm{i}c) \, \e^{\mathrm{i}(k_1x_1-k_2x_2)}\nonumber\\
& &-(k_1-k_2-\mathrm{i}c)(k_1+k_2-\mathrm{i}c) \, \e^{\mathrm{i}(-k_2x_1+k_1x_2)}\nonumber\\
& &+(k_1-k_2-\mathrm{i}c)(k_1+k_2+\mathrm{i}c) \, \e^{-\mathrm{i}(k_1x_1+k_2x_2)}\nonumber\\
& &+(k_1-k_2+\mathrm{i}c)(k_1+k_2+\mathrm{i}c) \, \e^{-\mathrm{i}(k_2x_1+k_1x_2)},
\label{WF-2}
\end{eqnarray}
provided that the Bethe equations
\begin{eqnarray}
\e^{\mathrm{i}2k_1L}&=&\frac{(k_1-k_2+\mathrm{i}c)(k_1+k_2+\mathrm{i}c)}{(k_1-k_2-\mathrm{i}c)(k_1+k_2-\mathrm{i}c)},\nonumber\\
\e^{\mathrm{i}2k_2L}&=&\frac{(k_2-k_1+\mathrm{i}c)(k_2+k_1+\mathrm{i}c)}{(k_2-k_1-\mathrm{i}c)(k_2+k_1-\mathrm{i}c)},
\end{eqnarray}
are satisfied.

Observe that for $c=0$, the wave function \eref{WF-2} reduces to the standing wave 
$\Psi(x_1,x_2)=\sin k_1x_1 \sin k_2x_2+\sin k_2x_1 \sin k_1x_2$,
with $k_1={n_1\pi}/{L}$ and $k_2={n_2\pi}/{L}$ for $n_1$ and $n_2$ non-zero integers. 
When $c$ increases, $k_1$ and $k_2$ also increase as if the boson mass increases. 
The standing wave properties are gradually lost as the interaction becomes stronger.

In a similar way, we can derive the $N$-particle wave function given in \eref{WF}. 
The coefficients are connected to each other via
\begin{equation}
A(\ldots, \epsilon_ik_i, \ldots, \epsilon_jk_j, \ldots)=\frac{\epsilon_ik_i-\epsilon_jk_j+\mathrm{i}c}{\epsilon_ik_i-\epsilon_jk_j-\mathrm{i}c}A(\ldots, \epsilon_jk_j, \ldots, \epsilon_ik_i, \ldots),
\end{equation}
for $i<j$. 
Application of the boundary conditions leads to the explicit form 
of the coefficients given in \eref{coff-A}, along with the Bethe equations \eref{BA}. 
The wave function is unnormalized.

\end{document}